\documentclass{article}
\usepackage[T1]{fontenc} 
\usepackage[utf8]{inputenc} 
\usepackage{ismir,amsmath,cite,url}
\usepackage{graphicx}
\usepackage{color}
\usepackage{booktabs}
\usepackage{tabularray}
\usepackage{multirow}
\usepackage{amsmath}
\usepackage{amssymb}
\usepackage{array}
\usepackage{makecell}
\usepackage{lineno}
\usepackage{xcolor}
\usepackage{microtype}

\title{Roman Numeral Analysis with Graph Neural Networks:
Onset-wise Predictions from Note-wise Features}

\oneauthor
  {Emmanouil Karystinaios$^1$ \hspace{1.5cm} Gerhard Widmer$^{1,2}$} {$^1$ Institute of Computational Perception, Johannes Kepler University Linz, Austria \\ $^2$ LIT AI Lab, Linz Institute of Technology, Austria\\ {\tt firstname.lastname@jku.at}}

\def\authorname{E. Karystinaios and G. Widmer}

\begin{document}

\maketitle

\begin{abstract}
Roman Numeral analysis is the important task of identifying chords and their functional context in pieces of tonal music. 
This paper presents a new approach to automatic Roman Numeral analysis in symbolic music. While existing techniques rely on an intermediate lossy representation of the score, we propose a new method based on Graph Neural Networks (GNNs) that enable the direct description and processing of each individual note in the score. 
The proposed architecture can leverage notewise features and interdependencies between notes but yield onset-wise representation by virtue of our novel edge contraction algorithm. 
Our results demonstrate that \textit{ChordGNN} outperforms existing state-of-the-art models, achieving higher accuracy in Roman Numeral analysis on the reference datasets. 
In addition, we investigate variants of our model using proposed techniques such as NADE, and post-processing of the chord predictions. The full source code for this work is available at \url{https://github.com/manoskary/chordgnn}

\end{abstract}

\section{Introduction}\label{sec:introduction}

Automatic Chord Recognition is one of the core problems in Music Information Retrieval. The task consists of identifying the harmonies or chords present in a musical piece. Various methods have been proposed to address this task using either an audio or symbolic representation of the music~\cite{pauwels201920}. In the symbolic domain, most approaches focus on the related and arguably more complex problem of Automatic Roman Numeral Analysis, which is a functional harmony analysis problem that has its roots in musicological research of Western classical music. 

Roman Numeral Analysis is a notational system used in music theory to analyze chord progressions and identify the relationship between chords in a given key. In this system, each chord in a piece of music is assigned a Roman numeral based on its position within the key's scale. For example, in the key of C major, the I chord is C major, the IV chord is F major, and the V chord is G major. Roman Numerals are an important tool for understanding and analyzing the harmonic structure of music, and they are a valuable resource for musicians, composers, and arrangers alike.

In Music Information Retrieval, a lot of work has been done to automate Roman Numeral analysis. However, current approaches still face significant challenges. Some of these are related to the large chord symbol vocabulary. A common way to address this problem is to divide a Roman Numeral into several components (e.g., key, degree, inversion) and transform the analysis into a multitask learning scenario. However, multitask approaches themselves face challenges with interdependencies among tasks. Lastly, Roman Numeral analysis faces a score representation problem related to existing models such as CNNs whose inputs must be in fixed-sized chunks. Recent state-of-the-art approaches follow an audio-inspired strategy, dividing a musical score into fixed-length time frames ("windows") which are then processed by a Convolutional Recurrent Neural Network (CRNN). However, such a representation is unnatural for scores and has the added practical disadvantage of being time-limited (for example regarding notes extending beyond the current window) and, due to the fixed-length (in terms of score time) constraint, capturing varying amounts of musically relevant context.

In this paper, we propose a new method for automatic Roman Numeral analysis based on Graph Neural Networks that can leverage note-wise information to address the score representation issue. Our model, \textit{ChordGNN}, builds on top of existing multitask approaches but introduces several novel aspects, including a graph convolutional architecture with an edge contraction pooling layer that combines convolution at the note level but yields the learned representation at the onset level.

Our proposed method, \textit{ChordGNN}, is evaluated on a large dataset of Western classical music, and the experimental results demonstrate that it outperforms existing state-of-the-art methods, in terms of the commonly used Chord Symbol Recall measure. To address the interdependencies among tasks we investigate the effect of post-processing and other proposed techniques such as NADE and gradient normalization.
Finally, we look at a qualitative musical example and compare our model's predictions with other state-of-the-art models.

\section{Related Work}\label{sec:related}

There is a big body of literature covering the topic of Automatic Chord Recognition applied in the audio domain; however, in our work, we focus on the problem of automatic Roman Numeral Analysis in the symbolic domain. It consists of labeling the chords and harmonic progressions in a piece of music using Roman Numerals, where each numeral represents a chord built on a particular scale degree. Numerous approaches have tried to automate Roman Numeral analysis or infer harmonic relations between chords. Notable work includes statistical models such as \textit{Melisma}~\cite{temperley2004cognition}, HMM-based models~\cite{raphael2004functional}, and grammar-based approaches~\cite{magalhaes2011functional}.

\begin{figure}[t]
    \centering
    \includegraphics[width=0.95\columnwidth]{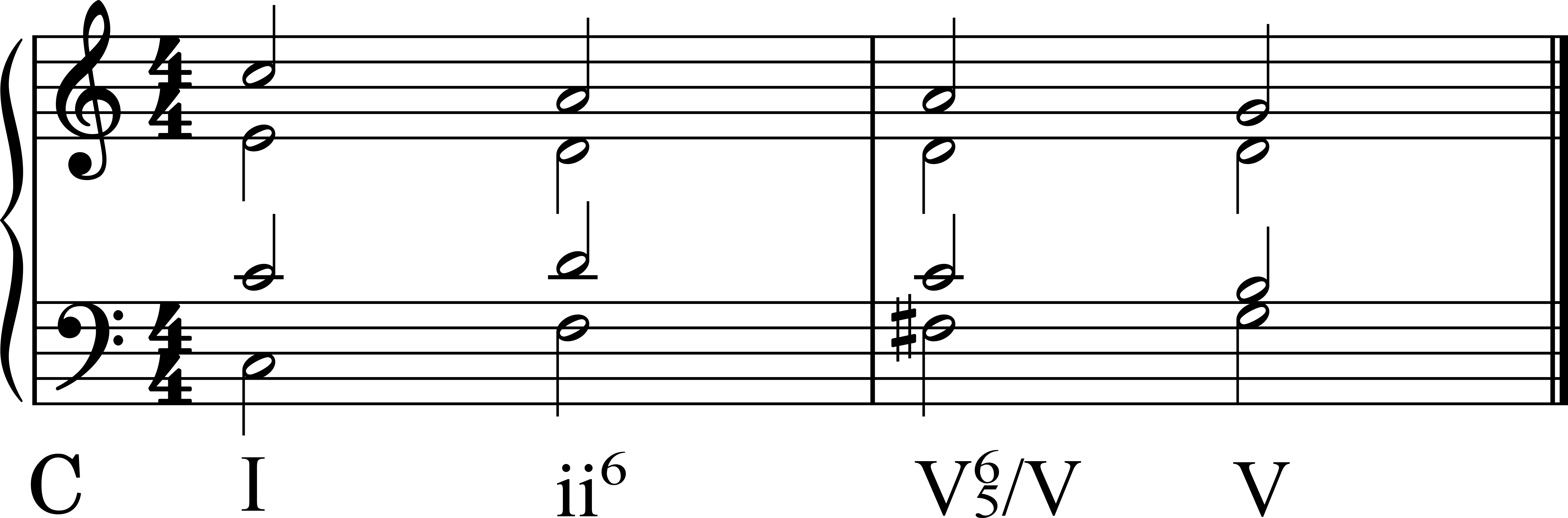}
    \caption{A Roman Numeral analysis for two bars for four-part harmony in $C$ major. Capital letters stand for major quality and lowercase for minor quality. The third chord has a dominant seven as its primary degree and the dominant of $C$ major as its secondary degree. The $V^6_5$ indicates a major with a seven quality in second inversion. The bass (lowest chord note) of that chord is $F$ sharp, the root is $D$, and the local key is $C$ major.    
    }
    \label{fig:rna_example}
\end{figure}

In recent years, research has shifted towards a deep learning and data-driven approach. Due to the large vocabulary of possible Roman Numerals, the problem has been divided into several component subtasks, thus resulting in a multitask learning setting~\cite{chen2018functional}. As a multitask problem, a Roman Numeral is characterized by the following components: the primary and secondary degree (as illustrated in Figure~\ref{fig:rna_example}), the local key at the time point of prediction, the root of the chord, the inversion of the chord, and the quality (such as major, minor, 7, etc.). Although the root can be derived from the other components, it was pointed out by  \cite{micchi2020not} that redundancy is assisting Roman Numeral analysis systems to learn. An example of Roman Numerals and their components can be viewed in Figure~\ref{fig:rna_example}. Recent state-of-the-art approaches decompose the numeral prediction task to the simultaneous prediction of those 6 components~\cite{chen2018functional, micchi2020not, lopez2021augmentednet, micchi2021deep, mcleod2021modular}.

Most deep learning approaches to Roman Numeral analysis are inspired by work in audio classification, cutting a score into fixed-size chunks (in terms of some constant score time unit; e.g., a 32nd note) and using these as input to deep models. Using this quantized time frame representation, \cite{micchi2020not} introduced a CRNN architecture to predict Roman Numerals. Other work has continued to build on the latter by introducing more tasks to improve performance such as the \textit{AugmentedNet} model~\cite{lopez2021augmentednet}, or introducing intra-dependent layers to inform in an orderly fashion the prediction of one task with the previously predicted task, such as the model introduced by \cite{micchi2021deep}. Other architectures, such as the CSM-T model, have demonstrated good results by introducing modular networks which treat a score as a sequence of notes ordered first by onset and then by pitch\cite{mcleod2021modular}.

\begin{figure}[t]
    \centering
    \includegraphics[width=0.75\columnwidth]{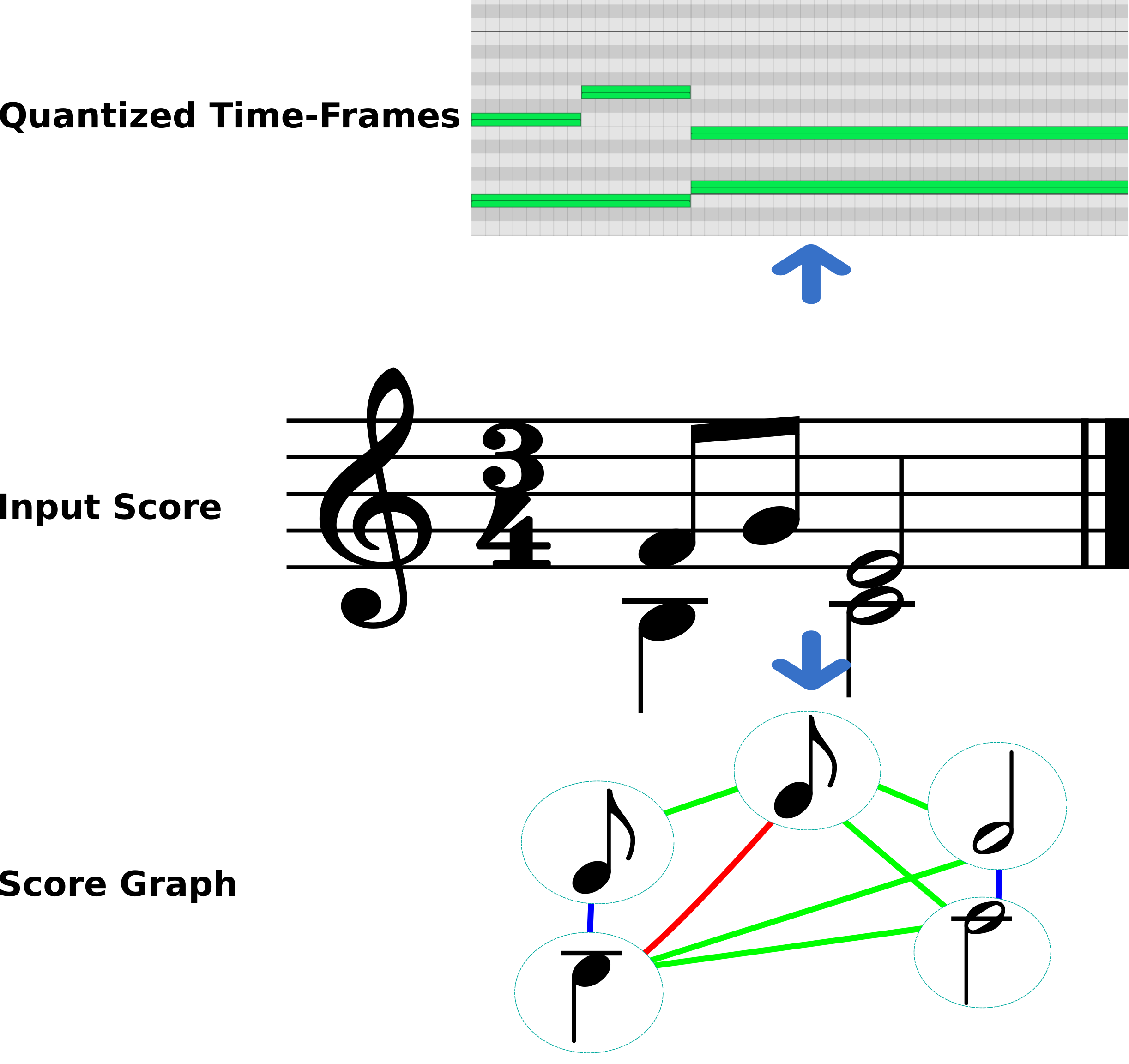}
    \caption{Different representations of the score excerpt shown in the middle. Top: quantized time frame representation, bottom: graph representation.}
    \label{fig:representations}
\end{figure}

Should a musicologist perform music analysis on a piece of music, they would consider the individual notes existing in the score. Thus, a time frame representation would come across as unnatural for symbolic music and in particular for such an analysis task. In this paper, we present a method that no longer treats the score as a series of quantized frames but rather as a partially ordered set of notes connected by the relations between them, i.e., a graph. A visual comparison of the two representations is shown in Figure~\ref{fig:representations}.
Recently, modeling scores as graphs has also been demonstrated to be beneficial for problems such as expressive performance generation~\cite{jeong2019graph}, cadence detection~\cite{karystinaios2022cadence}, voice separation~\cite{vocsep}, or boundary detection~\cite{hernandezolivan2023symbolic}.

Automatic Roman Numeral analysis, as a multitask problem, is mostly tackled with hard parameter-sharing models. These models share part of the model across all tasks as an encoder, and then the common embeddings are branched to a classification model per task~\cite{micchi2020not, lopez2021augmentednet, micchi2021deep}. However, some approaches separate tasks from this paradigm to a more modular or soft parameter sharing approach~\cite{mcleod2021modular}.

In the field of multitask learning, a lot of research has been done on the problem of conflicting gradients during backpropagation in hard parameter-sharing models. 
Issues with multi-objective optimization have been early addressed by Zhang et al.~\cite{zhang2014facial} and recent solutions have been proposed for the multitask setting in the form of dynamic task prioritization~\cite{guo2018dynamic}, gradient normalization~\cite{chen2018gradnorm}, rotation matrices~\cite{javaloy2022rotograd}, or even game-theoretic approaches~\cite{navon2022multi}. In our work, we experimentally evaluate some of these techniques in the multitask setting to investigate whether Roman Numeral analysis subtasks conflict with each other (see Section~\ref{subsec:configstudy}).

\section{Methodology}\label{sec:methodology}

\begin{figure*}[htbp]
    \centering
    \includegraphics[width=0.90\textwidth]{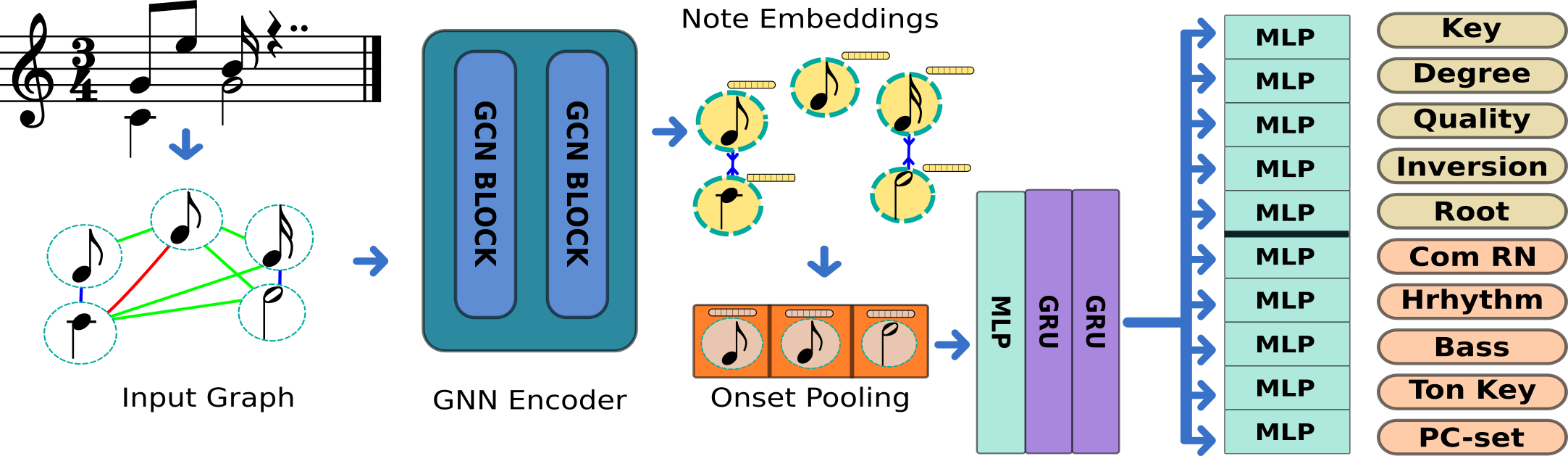}
    \caption{The proposed Architecture Chord-GNN}
    \label{fig:my_model}
\end{figure*}

\subsection{Roman Numeral Analysis}\label{subsec:rna}

We already discussed, in Section~\ref{sec:related}, how Roman Numeral analysis can be viewed as a multi-task problem. In this section, we describe in detail the additional tasks introduced by~\cite{lopez2021augmentednet} that we also use for training and prediction. First, let us assume that the prediction can be broken down into specific time points, and each time point is attributed to a unique onset in the score.

The Roman Numeral prediction can be viewed as a simultaneous prediction of the local key, degree (primary and secondary), quality, inversion, and root. Each one of these tasks is a categorical, multiclass classification problem. However, \cite{lopez2021augmentednet} indicated that only three tasks would be sufficient for $~98\%$ of the Roman Numeral annotations in our dataset (detailed in Section~\ref{subsec:datasets}). These three tasks comprise the prediction of a restricted vocabulary of common Roman Numeral symbols in combination with the local key and the inversion. We refer to Roman Numeral prediction involving the 5 tasks as \textit{conventional RN}, and the combined prediction of key, inversion, and restricted RN vocabulary \textit{alternative RN}, as $RN_{alt}$, in accordance with \cite{lopez2021augmentednet}.

Several other tasks have been introduced that have been shown to improve the performance of related models~\cite{lopez2021augmentednet}. These include the Harmonic Rhythm, which is used to infer the duration of a Roman Numeral at a given time point; the Tonicization task, a multiclass classification task that refers to a tonicized key implied by the Roman Numeral label and is complementary to the local key; the Pitch Class Sets task, which includes a vocabulary of different pitch class sets, and the Bass task, which aims to predict the lowest note in the Roman Numeral label.

\subsection{Graph Representation of Scores}\label{subsec:score}

Our approach to automatic Roman Numeral analysis no longer treats the score as a sequence of quantized time frames but rather as a graph, which permits us to specify note-wise information such as pitch spelling, duration, and metrical position. We use graph convolution to model interdependencies between notes. We model our score generally following Karystinaios and Widmer~\cite{karystinaios2022cadence}, but we opt for a heterogeneous graph convolution approach, i.e., including different edge relations/types. Furthermore, we develop an edge contraction pooling layer that learns onset-wise representations from the note-wise embeddings and therefore yields a sequence.

After the edge contraction, we follow ~\cite{micchi2020not, lopez2021augmentednet, micchi2021deep} by adding to the graph convolution a sequence model for the hard-sharing part of our model, and simple shallow multi-layer perceptron heads for each task. In essence, we replace the CNN encoder that works on quantized frames of the score in previous approaches, with a graph convolutional encoder followed by an edge contraction layer. Our proposed architecture is shown in Figure~\ref{fig:my_model}.

The input to the GNN encoder is an attributed graph $G = (V, E, X)$ where $V$ and $E$ denote its node and edge sets and $X$ represents the node feature matrix, which contains the features of the notes in the score. For our model, we used pitch spelling, note duration, and metrical position features.

Given a musical piece, the graph-building process creates a set of edges $E$, with different relation types $\mathcal{R}$. A labeled edge $(u, r, v)$ of type $r$ between two notes $u, v$ belongs to $E$ if the following conditions are met:
\begin{itemize}
    \item notes starting at the same time: \\ $on(u) = on(v) \xrightarrow{} r = \textrm{onset}$ 
    \item note starting while the other is sounding: $on(u) > on(v) \land on(u) \leq on(v)+dur(v) \xrightarrow{} r = \textrm{during}$ 
    \item note starting when the other ends: \\ $on(u) + dur(u) = on(v) \xrightarrow{} r = \textrm{follow}$
    \item note starting after a time frame when no note is sounding: $on(u) + dur(u) < on(v) \land \nexists v' \in V, \; on(v') < on(v) \land on(v') > on(u) + dur(u) \xrightarrow{} r = \textrm{silence}$
\end{itemize} 

\subsection{Model}\label{subsec:model}
In this section, we introduce and describe \textit{ChordGNN}, a Graph Convolutional and Recurrent Neural Network. The structure of the network is visually outlined in Figure~\ref{fig:my_model}. \textit{ChordGNN} uses heterogeneous graphSAGE~\cite{hamilton2017inductive} convolutional blocks defined as:
\begin{align}\begin{aligned}
    \mathbf{h}_{\mathcal{N}_r(v)}^{(l+1)} &= \mathrm{mean}\left(\{\mathbf{h}_{u}^{l}, \forall u \in \mathcal{N}_r(v) \}\right)\\
    \mathbf{h}_{v_r}^{(l+1)} &= \sigma \left(W \cdot \mathrm{concat}(\mathbf{h}_{v}^{l}, \mathbf{h}_{\mathcal{N}_r(v)}^{l+1}) \right)\\
    \mathbf{h}_{v}^{(l+1)} &= \frac{1}{|\mathcal{R}|} \sum_{r \in \mathcal{R}} \mathbf{h}_{v_r}^{(l+1)}
\end{aligned}\end{align}

where $\mathbf{h}_{v}^{(0)} = \mathbf{x}_v$ and $\mathbf{x}_u$ is the input features for node $u$, $\mathcal{N}(u)$ are the neighbors of node $u$, and $\sigma$ is a ReLU activation function. We name the output representations of all nodes after graphSAGE convolution $H = \{h^{(L)}_u \mid u \in V \}$ where $L$ is the total number of convolutional layers.

Given the hidden representation $H$ of all nodes, and onset edges $E_{\textrm{On}} = \{ (u, v) \mid on(u) = on(v) \}$, the onset edge contraction pooling is described by the following equations:
first, we update the hidden representation with a learned weight, $H' = H W^{(\textrm{cpool})}$. Subsequently we need to unify the representations for every node $u$, such that $\forall v \in \mathcal{N}_{\textrm{On}}(v), \; h^{(\textrm{cp})}_u = h^{(\textrm{cp})}_v$:
\begin{equation}
    h^{(\textrm{cp})}_u = h_u + \sum_{v \in \mathcal{N}_{\textrm{On}}(v)} h_v
\end{equation}
where, $h_u$ and $h_v$ belong to $H'$. Subsequently, we filter the vertices:
\begin{equation}
    V' = \{ v \in V | \; \forall u \in V, \; (v, u) \in E_{\textrm{On}} \implies u \notin V' \}
\end{equation}

Therefore, $H^{(cp)} = \{ h_u^{(cp)} \mid \forall u \in V' \}$ are the representations obtained. Sorting the representations by the onset on which they are attributed we obtain a sequence $S = [h_{u_1}^{(cp)}, h_{u_2}^{(cp)},\dots h_{u_k}^{(cp)}]$ such that $on(u_1) < on(u_2) < \dots < on(u_k)$.

The sequence $S$ is then passed through an MLP layer and 2 GRU layers. This concludes the hard-sharing part of our model. Thereafter, an MLP head is attached per task, as shown in Figure~\ref{fig:my_model}.

For training, we use the dynamically weighted loss introduced by~\cite{liebel2018weightedloss}.
The total loss $\mathcal{L}_{tot}$ of our network is calculated as a weighted sum of the individual losses for every task, where the weights are learned during training:

\begin{equation}
    \mathcal{L}_{\textrm{tot}} = \sum_{t \in \mathcal{T}} \mathcal{L}_{t} * \frac{1}{2\gamma_{t}^2} + \log(1 + \gamma_{t}^2)
\end{equation}

where $\mathcal{T}$ is the set of tasks; $\mathcal{L}_{t}$ is the cross-entropy loss relating to task $t$; the $\gamma_t$ are learned scalars that give the weight for each task $t$; and the $\log$ expression is a regularization term \cite{liebel2018weightedloss}.

\begin{figure}[b]
    \centering
    \includegraphics[width=\columnwidth]{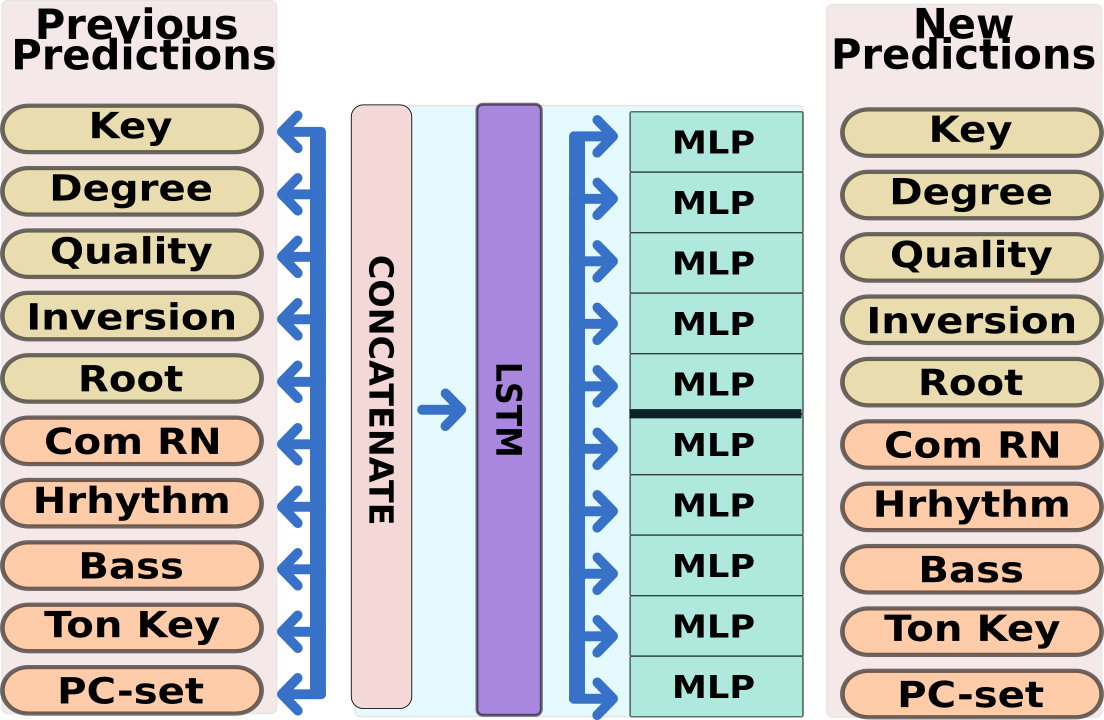}
    \caption{Post-processing of Roman Numeral predictions.}
    \label{fig:post-process}
\end{figure}

\subsubsection{Post-processing}\label{subsubsec:postprocess}

We enhance our model with a post-processing phase after the model has been trained. The post-processing phase combines the logits of all tasks' predictions by concatenating them and, then, feeds them to a single-layer bidirectional LTSM block. Then, again the embeddings of the sequential block are distributed to 11 one-layer MLPs, one for each task. The post-processing block is sketched in Figure~\ref{fig:post-process}.

\section{Experiments and Corpora}\label{sec:experiments}

In the experiments, we compare our model, \textit{ChordGNN}, with other recent models for automatic Roman Numeral analysis. We run experiments with our model in the exact same way as described in the paper~\cite{lopez2021augmentednet}, including the specific data splits, so that our results are directly comparable to the figures reported there.
A detailed comparison of the results will be given in Table~\ref{tab:full_comparison}. Furthermore, we develop variants of our model using proposed techniques such as NADE~\cite{micchi2021deep}, and post-processing of the chord predictions. We report a configuration study of our model on the use of gradient normalization techniques and NADE that should improve results on Multi-Task learning scenarios and avoid common Multi-Task Learning problems such as conflicting gradients. Lastly, we compare our model with the updated version \textit{v1.9.1} of the state-of-the-art model Augmented-Net~\cite{lopez2023arna} and datasets.

\subsection{Datasets}\label{subsec:datasets}

\begin{table*}[htbp]
    \centering
    \begin{tabular}{c | l|c c c c c | c c c}
    & \multicolumn{1}{c}{\textbf{Model}} & \multicolumn{1}{c}{Key} & \multicolumn{1}{c}{Degree} & \multicolumn{1}{c}{Quality} & \multicolumn{1}{c}{Inversion} & \multicolumn{1}{c}{Root} &
        \multicolumn{1}{c}{RN} & \multicolumn{1}{c}{RN (Onset)} & \multicolumn{1}{c}{$\textrm{RN}_{\textrm{alt}}$} \\
        \hline \hline
       \parbox[t]{2mm}{\multirow{5}{*}{\rotatebox[origin=c]{90}{\textbf{BPS}}}} & Micchi (2020) & 82.9 & 68.3 & 76.6 & 72.0 & - & 42.8 & - & - \\
       & CSM-T (2021) & 69.4 & - & - & - & 75.4 & 45.9 & - & - \\
       & $\textrm{AugNet}$ (2021) & \textbf{85.0} & \textbf{73.4} & \textbf{79.0} & 73.4 & \textbf{84.4} & 45.4 & - & 49.3\\
       & ChordGNN (Ours) & 79.9 & 71.1 & 74.8 & 75.7 & 82.3 & 46.2 & 46.6 &  48.6 \\
       & ChordGNN+Post (Ours) & 82.0 & 71.5 & 74.1 & \textbf{76.5} & 82.5 & \textbf{49.1} & \textbf{49.4} & \textbf{50.4} \\
        \hline 
        \parbox[t]{2mm}{\multirow{3}{*}{\rotatebox[origin=c]{90}{\textbf{Full}}}}& $\textrm{AugNet}$ (2021) & \textbf{82.9} & 67.0 & \textbf{79.7} & 78.8 & 83.0 & 46.4 & - & 51.5\\
        & ChordGNN (Ours) & 80.9 & 70.1 & 78.4 & 78.8 & 84.8 & 48.9 & 48.4 & 50.4 \\
        & ChordGNN+Post (Ours) & 81.3 & \textbf{71.4} & 78.4 & \textbf{80.3} & \textbf{84.9} &  \textbf{51.8} & \textbf{51.2} & \textbf{52.9}
    \end{tabular}
    \caption{Model comparison on two different test sets, the Beethoven Piano Sonatas (BPS), and the full test set. $RN$ stands for Roman Numeral, $RN_{alt}$ for the alternative Roman Numeral computations discussed in Section~\ref{subsec:rna}.
    $RN(Onset)$ refers to onset-wise prediction accuracy, all other scores use the CSR score (see Section~\ref{sec:results}).
    Note that model \textit{CSM-T} reports \textit{Mode} instead of \textit{Quality}.}
    \label{tab:full_comparison}
\end{table*}

For training and evaluation, we combined six data sources into a single "Full" Dataset of Roman Numeral annotations in accordance with \cite{lopez2021augmentednet}: the Annotated Beethoven Corpus (ABC)~\cite{neuwirth2018annotated}; the annotated Beethoven Piano Sonatas (BPS) dataset~\cite{chen2018functional}; the Haydn String Quartets dataset (HaydnSun)~\cite{lopez2017automatic}; the TAVERN dataset~\cite{devaney2015theme}; a part of the When-in-Rome (WiR) dataset~\cite{gotham2019romantext, gotham2021openscore}; and the Well-Tempered-Clavier (WTC) dataset~\cite{gotham2019romantext} which is also part of the WiR dataset. 

Training and test splits for the full dataset were also provided by \cite{lopez2021augmentednet}. It is worth noting that the BPS subset splits were already predefined in \cite{chen2018functional}. In total, approximately 300 pieces were used for training, and 56 pieces were used for testing, proportionally taken from all the different data sources. We draw a distinction for the BPS test set, which includes 32 Sonata first movements and for which we ran an additional experiment. The full test set also includes the 7 Beethoven piano sonatas.

In addition to the above datasets, 
we include data augmentations identical to the ones described in \cite{lopez2021augmentednet}: texturization and transposition. The texturization is based on a dataset augmentation technique introduced by~\cite{lopez2020harmonic}. The transposition augmentation boils down to transposing a score to all the keys that lie within a range of key signatures that have up to 7 flats or sharps. It should be noted that the augmentations are only applied in the training split.

For our last experiment (to be reported on in Section \ref{subsec:latest} below), we add additional data that were recently introduced by \cite{lopez2023arna}. The additional data include the annotated Mozart Piano Sonatas (MPS) dataset~\cite{hentschel2021annotated} for which we also applied the aforementioned augmentations.

\subsection{Configuration}\label{subsec:configuration}

For all our experiments, we train our network with the AdamW optimizer. We fix our architecture with a hidden size of $256$, a learning rate of $0.0015$, a weight decay of $0.005$, and a dropout of $0.5$ which is applied to each learning block of our architecture.

\section{Results}\label{sec:results}

As an evaluation metric, we use Chord Symbol Recall (CSR) \cite{harte2010towards} where for each piece, the proportion of time is collected during which the estimated label matches the ground truth label. We apply the CSR at the 32nd note granularity level, in accordance with \cite{micchi2020not, lopez2021augmentednet, mcleod2021modular}.

\subsection{Quantitative Results}

In the first experiment, which compares our \textit{ChordGNN} to existing state-of-the-art approaches, we evaluate the full dataset, but also the annotated Beethoven Piano Sonatas (BPS)~\cite{chen2018functional} subset, which many previous approaches had also used.
The results are shown in Table \ref{tab:full_comparison}.
We present the CSR scores (where they are applicable) for Local Key, Degree, Quality, Inversion, Root, conventional Roman Numeral, and Alternative Roman Numeral (see Section~\ref{sec:methodology}). Furthermore, we include the onset-wise accuracy score for our models' conventional Roman Numeral predictions.

On the BPS subset, we compare our model \textit{ChordGNN} with the Micchi (2020) model~\cite{micchi2020not}, the \textit{CSM-T} (2021) model~\cite{mcleod2021modular} and the \textit{AugmentedNet} 2021 model~\cite{lopez2021augmentednet}. Our results on Roman Numeral prediction surpass all previous approaches. Note that the \textit{AugmentedNet} model exhibits higher prediction scores on the individual Key, Degree, Quality, and Root tasks, which are used jointly for the prediction of the Roman numeral. These results indicate that our model obtains more meaningfully interrelated predictions, with respect to the Roman numeral prediction, resulting in a higher accuracy score.

Moreover, we compare \textit{ChordGNN} to \textit{AugmentedNet} on the full test dataset. Our model surpasses \textit{AugmentedNet} with and without post-processing in all fields apart from local key prediction and quality. Our model obtains up to $11.6\%$ improvement in conventional Roman Numeral prediction.

In both experiments, post-processing has been shown to improve both $RN$ and $RN_{alt}$. However, \textit{ChordGNN} without post-processing already surpasses the other models.

\begin{table}[t]
    \centering
    \begin{tabular}{l |c c}
        \textbf{Variant} & RN & $\textrm{RN}_{\textrm{alt}}$ \\
        \hline
         ChordGNN (Baseline) & $ 46.1 \pm 0.003$ & $ 47.8 \pm 0.007$\\
         ChordGNN + WLoss & $ \mathbf{48.9} \pm 0.001$ & $ \mathbf{50.4} \pm 0.010$\\
         ChordGNN + Rotograd & $ 45.5 \pm 0.003$ & $ 47.1 \pm 0.005$\\
         ChordGNN + R-GradN & $ 45.2 \pm 0.006$ & $ 46.7 \pm 0.005$\\
         ChordGNN + NADE & $ 48.2 \pm 0.005$ & $ 49.9 \pm 0.005$
    \end{tabular}
    \caption{Configuration Study: Chord Symbol Recall on Roman Numeral analysis on the full test set. $RN$ stands for Roman Numeral, $RN_{alt}$ refers to the alternative Roman Numeral computations discussed in section~\ref{subsec:rna}. 
    WLoss stands for the dynamically weighted loss described in Section~\ref{fig:my_model}, and R-GradN stands for Rotograd with Gradient Normalization.
    Every experiment is repeated 5 times with the same ChordGNN model as Table 1 without post-processing.}
    \label{tab:ablation}
\end{table}

\begin{figure*}[htbp!]
    \centering
    \includegraphics[width=\textwidth]{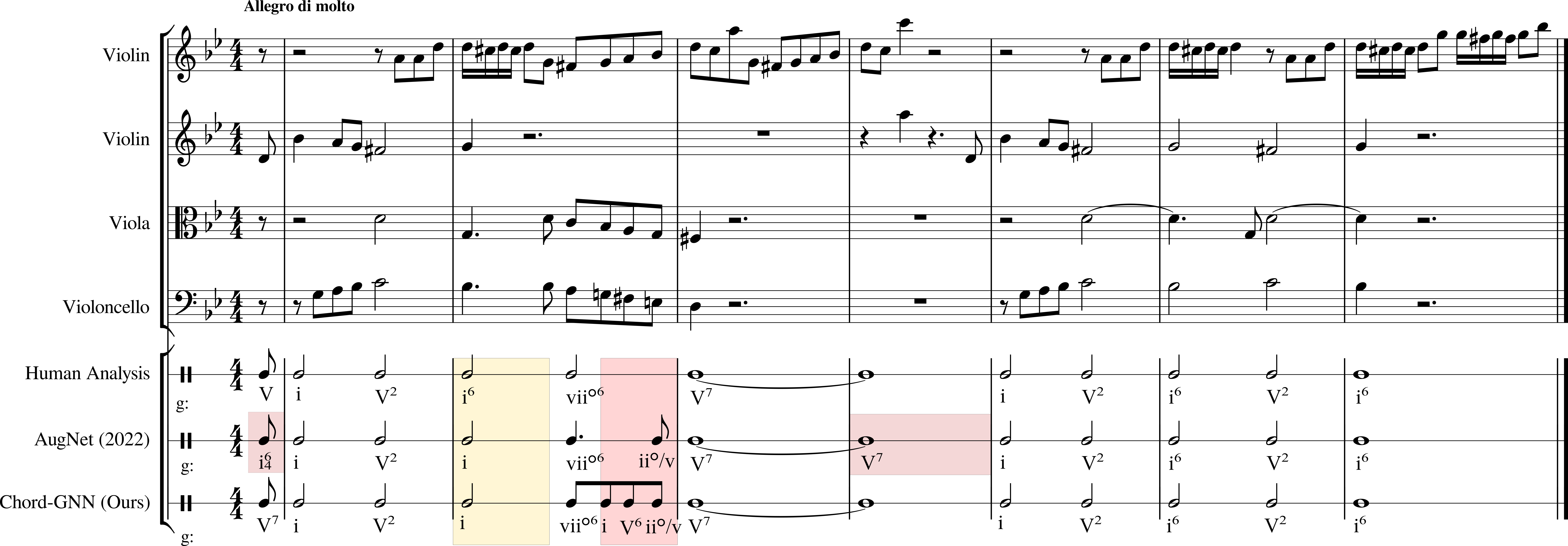}
    \caption{A comparison between the human annotation, AugmentedNet, and ChordGNN on a passage of Haydn's string quartet op.20 No.3 movement 4. The red (wrong) markings on Human Analysis and AugNet (2022) are from~\cite{lopez2023arna}}    
    \label{fig:arna_haydn_qualitative}
\end{figure*}

\subsection{Configuration Study}\label{subsec:configstudy}

For a systematic study of multitask training,
we investigated the effects of extension modules, gradient normalization techniques, and learnable weight loss. In detail, we test 5 configurations using as baseline the \textit{ChordGNN} model (without post-processing) with standard CE loss and no weighing. Furthermore, we test our proposed architecture using the dynamically weighted loss described in Section~\ref{subsec:model} (same as the model in Table~\ref{tab:full_comparison}), Rotograd~\cite{javaloy2022rotograd} and GradNorm~\cite{chen2018gradnorm} for Gradient Normalization, and NADE~\cite{micchi2021deep}. The models are run on the Full data set described above and averaged over five runs with random initialization. The results, summarized in Table \ref{tab:ablation}, suggest that using the dynamically weighted loss yields better results compared to other methods such as the Baseline or Gradient Normalization techniques. Furthermore, the dynamically weighted loss is comparable to NADE but also more robust on Conventional Roman Numeral prediction on our datasets.

\subsection{Latest developments}\label{subsec:latest}

Our last experiment focuses on specific developments that have very recently been published in N\'apoles L\'opez's Ph.D. thesis~\cite{lopez2023arna}. 
In the thesis, three additional tasks, related to predicting the components of a canonical representation of the current chord, as implied by the Roman Numeral, were proposed and the dataset was extended with the Annotated Mozart Piano Sonatas (MPS) corpus~\cite{hentschel2021annotated}, as mentioned in Section \ref{subsec:datasets} above.

To test the relevance of these updates, we trained an adapted version of our model, now with 11+3=14 individual tasks and including the Mozart data.
It turns out that the updated model improves significantly in performance, achieving a $53.5$ CSR score on conventional Roman Numeral (compare this to row "ChordGNN (Ours)" in Table \ref{tab:full_comparison}). Furthermore, post-processing can improve the results by up to two additional percentage points.\footnote{Unfortunately, we cannot directly compare these numbers to \cite{lopez2023arna}, as their results are not reported in comparable terms.}

\subsection{A Musical Example}\label{subsec:mus_example}

In Figure~\ref{fig:arna_haydn_qualitative}, we look at a comparison between the human annotations, \textit{AugmentedNet} and \textit{Chord-GNN} predictions (The musical excerpt is taken from N\'apoles L\'opez's thesis~\cite{lopez2023arna}, and the predictions relate to the new models trained as described in the previous section.).
Marked in red are false predictions, and marked in yellow are correct predictions of the model with wrong ground-truth annotations. Both models' predictions are very similar to the human analysis. However, our model correctly predicts the initial pickup measure annotation. In measure 2, the ground truth annotation marks a tonic in first inversion; however, the viola at that point is lower than the cello and therefore the chord is actually in root position. Both models obtain a correct prediction at that point. Subsequently, our model predicts a harmonic rhythm of eighth notes, which disagrees with the annotator's half-note marking. Analyzing the underlying harmony in that passage, we can justify our model's choices. 

The human annotation suggests that the entire second half of the 2nd measure represents a $vii^o$ chord. However, it should not be in the first inversion, as the cello plays an F\# as the lowest note (which is the root of $vii^o$). The AugNet analysis faces the same issue, in contrast with the predictions of ChordGNN. However, there are two conflicting interpretations of the segment. First, the $vii^o$ on the third beat is seen as a passing chord between the surrounding tonic chords, leading to a dominant chord in the next measure. Alternatively, the $vii^o$ could already be part of a prolonged dominant harmony (with passing chords on the offbeats) leading to the $V^7$. The ChordGNN solution accommodates both interpretations as it doesn't attempt to group chords at a higher level, treating each eighth note as an individual chord rather than a passing event. The other two solutions prefer the second option.

\section{Conclusion}\label{sec:conclusion}

In this paper, we presented \textit{ChordGNN}, a model for automatic Roman Numeral analysis in symbolic music, based on a note-level, graph-based score representation. We showed that \textit{ChordGNN} improves on other state-of-the-art models, and that post-processing can further improve the accuracy of the predictions. A configuration study suggests that gradient normalization techniques or techniques for carrying prediction information across tasks are not particularly beneficial or necessary for such a model.

Follow-up work will focus on strengthening the robustness of our models by pre-training with self-supervised methods on large corpora. We believe that such pre-training can be beneficial for learning helpful intrinsic musical information. Such a step is crucial since more data improves predictions but Roman Numeral annotations are hard to find or produce. Moreover, we aim to enrich the number of tasks for joint prediction by including higher-level analytical targets such as cadence detection and phrase boundary detection. Finally, we aim to extend our method to the audio domain.

\section{Acknowledgements}

We gratefully acknowledge the musical analysis of the $vii^o$ passage in Fig.~\ref{fig:arna_haydn_qualitative} (Section ~\ref{subsec:mus_example}) that was offered by an anonymous reviewer, and which we took the liberty of adopting for our text. This work is supported by the European Research Council (ERC) under the EU’s Horizon 2020 research \& innovation programme, grant agreement No.~101019375 (“Whither Music?”), and the Federal State of Upper Austria (LIT AI Lab).

\bibliography{biblio}

\end{document}